\documentclass[twocolumn,aps,prb]{revtex4}
\usepackage{amsmath}

\begin{document}

\title{Reply to 'Comment on "Dynamic correlations of the spinless Coulomb
Luttinger liquid [Phys. Rev. B {\bf 65}, 125109 (2002)]"'}

\author{Yasha~Gindikin and V.~A.~Sablikov}

\affiliation{Institute of Radio Engineering and Electronics,
Russian Academy of Sciences, Fryazino, Moscow District, 141120,
Russia}

\begin{abstract}
We show that the criticism of our paper [Phys. Rev. B {\bf 65},
125109 (2002)] by Wang, Millis, and Das Sarma [cond-mat/0206203]
is based on a trivial mathematical mistake they have committed.
\end{abstract}

\maketitle

Coulomb interaction in one-dimensional electron systems arouses
firm and enduring interest for two reasons. First, this is a
strong interaction of clear and fundamental nature. Second, the
Coulomb interaction case is usually difficult to treat
mathematically, and therefore exact solutions are always
important.

We have investigated the Luttinger liquid (LL) with Coulomb
electron-electron interaction in our paper~[\onlinecite{GS}]. The
main result of our work is apparently the existence of a soft
collective charge mode, related to dynamic electron correlations
of the $2k_F$ scale. Ref.~[\onlinecite{GS}] presents also an
analytic method to study various dynamic correlation functions
near the threshold.

Wang, Millis, and Das Sarma (WMS henceforth) claim in their
Comment~[\onlinecite{wang02}] that our method, as well as ensuing
results, is incorrect. The claim is based on WMS's abortive
attempt to reproduce our calculation of the CDW structure factor.
WMS substitute our expression for the structure factor into the
integral equation, which we derived for it, and conclude that
corrections to our result are given by diverging integrals. After
that, WMS accuse our method to be 'inconsistent with mathematical
analysis'.

In this Reply, we show that WMS have committed a trivial
mathematical mistake in their calculations. Diverging integrals
appear in Ref.~[\onlinecite{wang02}] just because WMS have
incorrectly differentiated the structure factor.

Recall that the CDW structure factor $S(q,\omega)$ contains the
Heaviside function $\theta(\omega-\omega_q)$ as a factor, which
reflects the existence of the threshold.\cite{LP} We stress that
$\theta(\omega-\omega_q)$ depends on $\omega$ and $q$, and hence
must be differentiated when finding the derivative of
$S(q,\omega)$. WMS disregard this fact and differentiate
$S(q,\omega)$, ignoring $\theta$-function.  Below we demonstrate
that if one acts correctly, then all the integrals of our
paper~[\onlinecite{GS}] are well-defined, and expansions are
convergent.

In Ref.~[\onlinecite{GS}] we have shown that $S(q,\omega)$
satisfies the following integral equation:
\begin{equation}
\dfrac{\omega}{v_F}S(q,\omega)=
\int_{-\infty}^{+\infty}dQ\,S(Q,\omega-\omega_{q+Q}).
\label{preexp}
\end{equation}
Here $\omega_q$ is the energy of the LL bosons,
$\omega_q=v_F|q|/g(q)$, $v_F$ is the Fermi velocity, $g(q)$ is the
interaction parameter.\cite{Voit} For Coulomb interaction
$g(q)=\beta |\ln |q|d|^{-1/2}$, with $d$ being the diameter of a
quantum wire, $\beta=[\pi \hbar v_F/2e^2]^{1/2}$. For simplicity,
the wave number $q$ is measured from $2k_F$.

Expand $S(Q,\omega-\omega_{q+Q})$ in the powers of the wave number
$Q$, appearing in the second (frequency) argument, to get
\begin{equation}
\begin{array}{l}
\displaystyle
\dfrac{\omega}{v_F}S(q,\omega)=\int_{-\infty}^{+\infty}dQ\,[S(Q,\omega-
\omega_{q})\\
\displaystyle
+\dfrac{Q^2}{2!}S_{qq}(Q,\omega-\omega_q)+\dots].
\end{array}
\label{expansion}
\end{equation}
Restricting expansion~(\ref{expansion}) to the first term on the
RHS, we have found in Ref.~[\onlinecite{GS}] that
\begin{equation}
S(q, \omega)= \dfrac{v_F}{\omega}\dfrac{e^{-4\beta|\ln
\epsilon|^{1/2}}}{\epsilon |\ln \epsilon|^{1/2}}\theta(\epsilon),
\label{SF}
\end{equation}
where we denoted $\epsilon=\omega-\omega_q$. The structure factor
$S(q,\omega)$ is seen to be zero for $\epsilon<0$, and to diverge
as $\epsilon \to +0$.

In our paper~[\onlinecite{GS}] we have emphasized that the Coulomb
interaction case is special in the respect that retaining only the
first term on the RHS of Eq.~(\ref{expansion}) already gives the
correct asymptotic behavior of $S(q,\omega)$ at $\epsilon \to 0$.

On the contrary, WMS find in Ref.~[\onlinecite{wang02}] that the
second term on the RHS of Eq.~(\ref{expansion}) is infinite, and
conclude that the expansion~(\ref{expansion}) does not exist at
all.

Let us prove that the second term of the
expansion~(\ref{expansion}) is well-defined, and corrections to
our result~(\ref{SF}) are indeed small. Substitute $S(q,\omega)$
from Eq.~(\ref{SF}) into Eq.~(\ref{expansion}), as WMS
wish~[\onlinecite{wang02}], but do it correctly at this time. The
second term of the expansion becomes
\begin{equation}
\begin{array}{l}
\displaystyle
\partial_{qq}\int_{-\infty}^{+\infty}dQ\,\dfrac{Q^2}{2!}
S(Q,\epsilon)\\
\displaystyle
=\partial_{qq}\int_0^{+\infty}dQ\,Q^2
\dfrac{v_F}{\epsilon} \left[\dfrac{e^{-4\beta|\ln
\delta|^{1/2}}}{\delta |\ln \delta|^{1/2}}\right]\theta(\delta),
\end{array}
\label{second}
\end{equation}
where we denoted $\delta=\epsilon-\omega_Q$. It is important that
both $\epsilon$ and $\delta$ depend on $q$, which must be taken
into account when calculating the derivative $\partial_{qq}$. When
$\epsilon<0$, $\delta$ is always negative, and
expression~(\ref{second}) is zero due to the factor
$\theta(\delta)$. If $\epsilon>0$, then $\delta$ varies from
$\epsilon$ to 0. Take $\delta$ as a new integration variable.
Eq.~(\ref{second}) can be written as
\begin{equation}
\partial_{qq}\int_0^{\epsilon}d\delta\,
\dfrac{\beta^3}{v_F^2}\dfrac{(\epsilon-\delta)^2}{\epsilon|\ln
(\epsilon-\delta)|^{3/2}}
 \left[\dfrac{e^{-4\beta|\ln
\delta|^{1/2}}}{\delta |\ln \delta|^{1/2}}\right]\theta(\epsilon).
\label{int}
\end{equation}
As WMS note~[\onlinecite{wang02}], the most important contribution
to the integral comes from the region $\delta \sim 0$, where the
expression in the square brackets of Eq.~(\ref{int}) diverges. For
this reason we can replace all $(\epsilon-\delta)$ in the integral
with~$\epsilon$. Since the expression in the square brackets
equals identically
$(2\beta)^{-1}\partial_{\delta}(\exp(-4\beta|\ln \delta|^{1/2}))$,
the total formula~(\ref{int}) is
\begin{equation}
\partial_{qq}\left[\frac{\beta^2}{2v_F^2}\epsilon \,\frac{e^{-4\beta|\ln
\epsilon|^{1/2}}}{|\ln \epsilon|^{3/2}}\theta(\epsilon)\right].
\label{fin}
\end{equation}
We underline that this expression is a well-defined distribution,
rather than a classic function. Distributions normally arise when
expanding a function with a threshold into Taylor's series. Hence
it is not surprising that the higher order terms of the expansion
diverge. Simply they are not 'corrections' as WMS suppose. In
calculating corrections to our result, one cannot retain only two
terms of the expansion~(\ref{expansion}). The correct
procedure~[\onlinecite{Gelfand}] requires summing up the total
series in Eq.~(\ref{expansion}) to get the result in terms of
classic functions. We have performed such calculation, and have
not included the result into our paper~[\onlinecite{GS}], since it
consists in a not too important replacement of the argument in the
expression~(\ref{SF}) for $S(q,\omega)$. The argument
$\epsilon=\omega-\omega_q$ should be shifted by
$\omega'_q\epsilon/|\ln \epsilon|^{1/2}$, which shift is obviously
negligible as $\epsilon \to 0$.

Thus we confirm our result that Eq.~(\ref{SF}) gives the
asymptotic behavior of $S(q,\omega)$ as $\epsilon \to +0$. WMS's
claim~[\onlinecite{wang02}] that the true diverging behavior
should not be like Eq.~(\ref{SF}) is misleading and incorrect.

Now consider the {\it mistake}, which is the basis of the WMS
Comment~[\onlinecite{wang02}]. In estimating the
integral~(\ref{second}), WMS first calculate the derivative of the
integrand w.r.t.~$q$. This way of calculation, though less
economical than the one presented above, would nevertheless lead
them to correct result, provided that WMS calculate the derivative
correctly. They do not take into account that the argument
$\delta=\omega-\omega_q-\omega_Q$ of the $\theta(\delta)$-function
depends on $q$, using implicitly the incorrect relation
$$
\dfrac{\partial^2}{\partial^2 q}\left[\dfrac{e^{-4\beta|\ln
\delta|^{1/2}}}{\epsilon\delta |\ln
\delta|^{1/2}}\theta(\delta)\right]=\theta(\delta)\dfrac{\partial^2}{\partial^2
q}\left[\dfrac{e^{-4\beta|\ln \delta|^{1/2}}}{\epsilon\delta |\ln
\delta|^{1/2}}\right],
$$
which finally leads them to conclusion that the
integral~(\ref{second}) diverges (see the showy, yet erroneous
Eq.~(14) of Ref.~[\onlinecite{wang02}]). In regards to WMS's
appeal for the purity of the mathematical analysis, there is
nothing left but to refer them to the important work of
Leibniz~[\onlinecite{Leibniz}], which explains how the derivative
of the function product should be found.

\end{document}